\pdfoutput=1
\documentclass{LLNS/llncs}
\usepackage{makeidx} 
\usepackage{cite}
\usepackage{latexsym}
\usepackage{amsmath}
\usepackage{amssymb}
\usepackage{graphicx}
\usepackage{algorithm}
\usepackage{algorithmic}
\usepackage{multirow}
\usepackage{verbatim}

\newcommand{\mathsym}[1]{{}}

\graphicspath{{.}}

\begin{document}

\title{Visualizing Graphs with Node and Edge Labels}

\author{Yifan Hu\thanks{e-mail: yifanhu@research.att.com}}
\institute{AT\&T Labs, Shannon Laboratory, 180 Park Ave., Florham Park, NJ 07932.}

\maketitle

\begin{abstract}
When drawing graphs whose edges and nodes contain text or graphics, such information needs to be displayed without overlaps, either as part of
the initial layout or as a post-processing step. The core problem
in removing overlaps lies in retaining the structural
information inherent in a layout, minimizing the additional
area required, and keeping edges as straight as possible. 
This paper presents a unified node and edge overlap
removal algorithm that does well at solving this problem.

\end{abstract}
\section{Introduction}
Most existing graph layout algorithms for undirected graph treat nodes as points and edges as straight lines. In
practice, both nodes and edges often contain labels or graphics that need to be
displayed. Naively incorporating these can lead to labels that overlap (see, e.g., Figure~\ref{olympic_org}),
causing the information of some labels to occlude that of others. It is therefore important to 
remove such overlaps either by taking into account label sizes during layout or
as a post processing step. Assuming that the original layout captures significant structural
information such as clusters, the goal of any layout that avoids
overlaps should be to retain the ``shape'' of the layout based on
point nodes. 

The simplest is to scale up the drawing \cite{DBLP:journals/constraints/MarriottSTH03} while
preserving label size until the labels no longer overlap. This has
the advantage of preserving the shape of the layout,
but is often impractical due to inconveniently large drawings.
Label overlap removal is typically a trade-off between preserving the shape, 
limiting the area, and keep edges straight, with scaling at one extreme. 

\subsection{Related work: node label overlap removal}

Many techniques to avoid label overlaps have been devised. 
One approach is to make the node size part of the model of the layout
algorithm. For hierarchical layouts, the node size can be naturally
incorporated into the algorithm
\cite{Friedrich_Schreiber_2004,Sugiyama}.  For symmetric layouts,
various authors
\cite{Chang_Lin_Yen_03,Harel_Koren_2002,Li_Eades_Nikolov_05,Wang_Miyamoto_95} have extended the spring-electrical model
\cite{Eades_1984,Fruchterman_Reingold_1991} to take into account node
sizes, usually as increased repulsive forces. Node overlap removal
can also be built into the stress model \cite{Kamada_Kawai_1989} by
specifying the ideal edge length to avoid overlaps along the
graph edges.  Such heuristics, however, do not always remove
all of the node overlaps, and have to be augmented with a
post-processing step. 

An alternative approach is to remove node overlaps as a 
post-processing step after the graph is layed-out. Here the trade-off
between layout size and preserving the graph's shape is much more
explicit. A number of algorithms have been proposed.  For example, the Voronoi cluster
busting algorithm \cite{Gansner_North_1998,Lyons_Meijer_Rappaport_1998} works by iteratively forming a 
Voronoi diagram from
the current layout and moving each node to the center of its Voronoi
cell until no overlaps remain. The idea is that 
restricting each node to its corresponding Voronoi cell should
preserve the relative positions of the nodes. In practice,
because of the number of iterations often required and the use of
a rectangular bounding box (the latter might be improved by
using something like $\alpha$-shapes \cite{alpha}
to provide a more accurate boundary),
the nodes in the final drawing can be homogeneously distributed,
bearing little resemblance to the original layout.

Another group of post-processing algorithms is based on
maintaining the orthogonal ordering \cite{DBLP:journals/vlc/MisueELS95}
of the initial layout as a way to preserve its shape.
A force scan algorithm and variants were proposed
\cite{Hayashi_Inoue_Masuzawa_Fujiwara_1998,huang_lai_03,Li_Eades_Nikolov_05,DBLP:journals/vlc/MisueELS95} based on these constraints.
Marriott 
et al. \cite{Dwyer_Marriott_Stuckey_2005,DBLP:journals/constraints/MarriottSTH03}
presented a quadratic programming algorithm
which removes node overlaps while minimizing node displacement and
keeping the orthogonal ordering. Each of these algorithms has a separate pass in both the
horizontal and vertical directions. In practice, this
asymmetry often results in a layout with a distorted aspect ratio \cite{Gansner_Hu_2008}.

While preserving orthogonal ordering is certainly important, it is more general to preserve
relative proximity relations \cite{Hu_Koren_2008}. Gansner and Hu \cite{Gansner_Hu_2008} proposed 
a node overlap removal algorithm that used a proximity graph as a scaffolding to
maintain the proximity relations, and employed a sparse stress model to remove node overlaps.
The algorithm was demonstrated to be scalable for large graphs. It produces layout that closely resembles the
original layout, but is free of node overlaps, and takes comparatively little additional area.

\subsection{Related work: edge label overlap removal or placement}

Edge label overlap removal or placement has also been studied by many authors. 
One of the often studied problems, though not the focus of this paper, is that of edge label placement (ELP),
where the geometry of the graph, other than the labels, is assumed to be
fixed. The problem is finding the best placement positions for the
edge labels to minimize label overlaps, as well as making the
association between edges and their corresponding labels
unambiguous. This is highly related to the map labeling problem
\cite{Christensen_Marks_Shieber_1996,Kakoulis_Tollis_1998,Nascimento_Eades_2008}.  Kakoulis and Tollis
\cite{Kakoulis_Tollis_1996,Kakoulis_Tollis_1997,Kakoulis_Tollis_2001} proved that the ELP problem is NP hard, and
presented an algorithm for labeling the edges of hierarchical
drawings. The algorithm was also used for graph drawings of other
styles \cite{Dogrusoz_Kakoulis_Madden_Tollis_1998,Dogrusoz_Kakoulis_Madden_Tollis_2007}.

A number of algorithms were proposed in drawing graphs without
edge label overlaps.  Castell et al. \cite{Castell_Mili_Tollis_2001}
presented a procedure for drawing state charts, with node and edge
labels.  Edge label overlaps are avoided as part of the layout process
by truncating long label strings and allocating enough space
vertically between layers. Binucci et
al. \cite{Binucci_Didimo_Liotta_Nonato_2005} present MILP models to
compute optimal drawings without edge label overlaps, and an exact
algorithm and several heuristics to compute such drawings with minimum
area.  Klau and Mutzel \cite{Klau_Mutzel_1999} proposed a branch and
cut algorithm which computes optimally labeled orthogonal drawings.

In this paper we study the problem of visualizing undirected graphs with both node and edge labels.
We assume that an aesthetic, symmetric layout based on
point nodes is available. The task is to adjust the node position so that node labels and
edge labels can be placed without overlaps. In doing so, we want to retain the ``shape''
of the layout. To minimize the additional area required, we allow 
edges to be bent, but strive to draw them as straight as possible.
We present (Section~\ref{sec_algorithm}) a
unified overlap removal algorithm based on a proximity graph of the
position of node and edge labels in the original layout. Using this graph as a guide, it iteratively moves
the labels to remove overlaps, while keeping the 
relative positions between them as close to
those in the original layout as possible, and edges as straight as possible. The algorithm is similar to
the stress model \cite{Kamada_Kawai_1989} used for graph layout,
except that the stress function involves only a sparse selection of
all possible node pairs. It is an extension to the node overlap removal algorithm of
Gansner and Hu \cite{Gansner_Hu_2008} to deal with both node and edge labels.
We evaluate our algorithm on two
graphs from applications. Finally, Section~\ref{sec_conclusion}
presents a summary and topics for further study.

\section{Background\label{sec_notations}}

We use \(G = (V, E)\) to denote an undirected graph, with \(V\) the
set of nodes (vertices) and \(E\) edges. We use \(|V|\) and \(|E|\)
for the number of vertices and edges, respectively.  We let \(x_i\)
represent the current coordinates of vertex \(i\) in Euclidean space.

The aim of graph drawing is to find \(x_i\) for all \(i\in V\) so that
the resulting drawing gives a good visual representation of the
information in the graph. Two popular methods, the spring-electrical
model \cite{Eades_1984,Fruchterman_Reingold_1991}, and the stress
model \cite{Kamada_Kawai_1989}, both convert the problem of finding an
optimal layout to that of finding a minimal energy configuration of a
physical system. We describe here the stress model in more detail, as
we shall use a similar model for the purpose of label overlap removal
in Section~\ref{sec_algorithm}.

The stress model assumes that there are springs connecting all nodes
of the graph, with the ideal spring length equal to the graph
theoretical distance between nodes. The energy of this spring system
is

\begin{equation}
\sum _{i\neq j} w_{\text{ij}} \left(\left\|x_i - x_j\right\| - d_{\text{ij}}\right){}^2,\label{spring_model}
\end{equation}
where \(d_{\text{ij}}\) is the graph theoretical distance between
vertices \(i\) and \(j\), and \(w_{\text{ij}}\) is a weight factor,
typically $1/d_{\text{ij}}{}^2$. The layout that
minimizes the above stress energy is an optimal layout of the graph. A robust technique to find an optimal
of this model is stress majorization, where
the cost function (\ref{spring_model}) is bounded by a series of
quadratic functions from above, and the process of finding an optimum
becomes that of solving a series of linear systems
 \cite{Gansner_Koren_North_2005b}. 

In the stress model, the graph theoretical distance between all pairs
of vertices has to be calculated, leading to quadratic complexity 
in the number of vertices. There have been 
attempts (e.g.,  \cite{Brandes_Pich_2008,Gansner_Koren_North_2005b})
to simplify the stress function by considering only a
sparse portion of the graph.
Our experience, however, is that these 
techniques can fail to yield good layouts on real-life graphs. Therefore, algorithms
based on a spring-electrical model employing a multilevel
approach and an efficient approximation scheme for long range
repulsive forces  \cite{Hachul_Junger_2004,Hu_2005,Walshaw_2003} are still 
the most efficient choices when laying out
large graphs without consideration of the node or edge labels. It is worth pointing out that
the label overlap removal algorithm proposed in this paper works with 
any symmetric layout, regardless of how it is generated.

\section{A Unified Model for Label Overlap Removal\label{sec_algorithm}}

Our goal is to remove label overlaps while preserving the shape of the
initial layout by maintaining the proximity relations \cite{Hu_Koren_2008}
among the nodes. Recently we proposed an efficient algorithm PRISM \cite{Gansner_Hu_2008} to
remove node overlaps based on the {\it proximity stress model}. We first summarize this model, 
then describe how to extend it to
solve the more general problem of overlap removal for both node and edge labels.

\subsection{The PRISM algorithm for node overlap removal}

We first set up a rigid ``scaffolding'' structure so that while
vertices can move around, their relative positions are
maintained. This scaffolding is constructed using an approximation to
proximity graph \cite{Jaromczyk_Toussaint_1992}, the Delaunay
triangulation (DT).

We  then check every edge in DT and see if there are
any node overlap along that edge. 
Let $W_i$ and $H_i$ denote the half width and height
of the node $i$, and $x^0_i(1)$ and $x^0_i(2)$ the current X and Y coordinates of this node.
If $i$ and $j$ form an edge in the DT, we calculate 
the {\em overlap factor} of these two nodes,

\begin{equation}
t_{\text{ij}} = \text{max}\left(
\text{min}
   \left(
     \frac{W_i+W_j}{|x^0_i(1)-x^0_j(1)|},
     \frac{H_i+H_j}{|x^0_i(2)-x^0_j(2)|}
   \right)
, 1\right).\label{ofactor}
\end{equation}

\noindent For nodes that do not overlap, $t_{\text{ij}} = 1$. For
nodes that do overlap, such overlaps can be removed if we expand the
edge by this factor. Therefore we want to generate a layout such that an edge 
in the proximity graph has the ideal edge length close to $t_{\text{ij}} \|x^0_i -
x^0_j\|$. In other words, we want to minimize the following stress function

\begin{equation}
\sum _{(i,j) \in  E_{P}} w_{\text{ij}}\left(\left\|x_i - x_j\| - d_{\text{ij}}\right.\right){}^2\label{PRISM}
\end{equation} 

\noindent Here $d_{\text{ij}} = s_{\text{ij}}\|x^0_i - x^0_j\|$ is the ideal distance
for the edge $\{i,j\}$, $s_{\text{ij}}$ is a scaling factor related to the overlap factor $t_{\text{ij}}$ (see (\ref{smax})), $w_{\text{ij}} = 1/d_{\text{ij}}^2$ is 
a scaling factor, and $E_{P}$ is the set of edges of the proximity graph. We call (\ref{PRISM}) the {\em proximity stress model}.

Because DT is a planar graph, which has no more than \(3|V|-3\) edges,
the above stress function has no more than \(3|V|-3\)
terms.  Furthermore, because DT is rigid, it provides a good
scaffolding that constrains the relative position of the vertices and
helps to preserve the global structure of the original layout.

It is important that we do not attempt to remove overlaps in one iteration
by minimizing the above model with $s_{\text{ij}}=t_{\text{ij}}$ to avoid 
high localized stress causing deviation to the original layout, hence we damp the overlap factor by setting

\begin{equation} s_{\text{ij}} = \text{min}(t_{\text{ij}}, s_\text{max})\label{smax}\end{equation}
\noindent and try to remove overlaps a little at
a time. Here $s_\text{max} > 1$ is a number limiting the amount of overlap
we are allowed to remove in one iteration. Typically $s_\text{max} = 1.5$.

After minimizing (\ref{PRISM}), we arrive at a layout
that may still have node overlaps. We then regenerate the proximity graph using DT and
calculate the overlap
factor along the edges of this graph, and redo the minimization. This forms an iterative process
that ends when there are no more overlaps along the edges of the 
proximity graph. At this stage there may still be node overlaps between nodes that do
not constitute an edge of the DT, so a scan-line algorithm \cite{Dwyer_Marriott_Stuckey_2005}
is used to find all overlaps, and the proximity graph is augmented with additional edges,
where each edge consists of a pair of nodes that overlap.
We then re-solve 
(\ref{PRISM}). This process is repeated until the scan-line algorithm finds
no more overlaps. We called the algorithm PRISM (PRoxImity Stress Model).

\begin{figure}[htb]
\begin{center}
  \includegraphics*[width=5.2in]{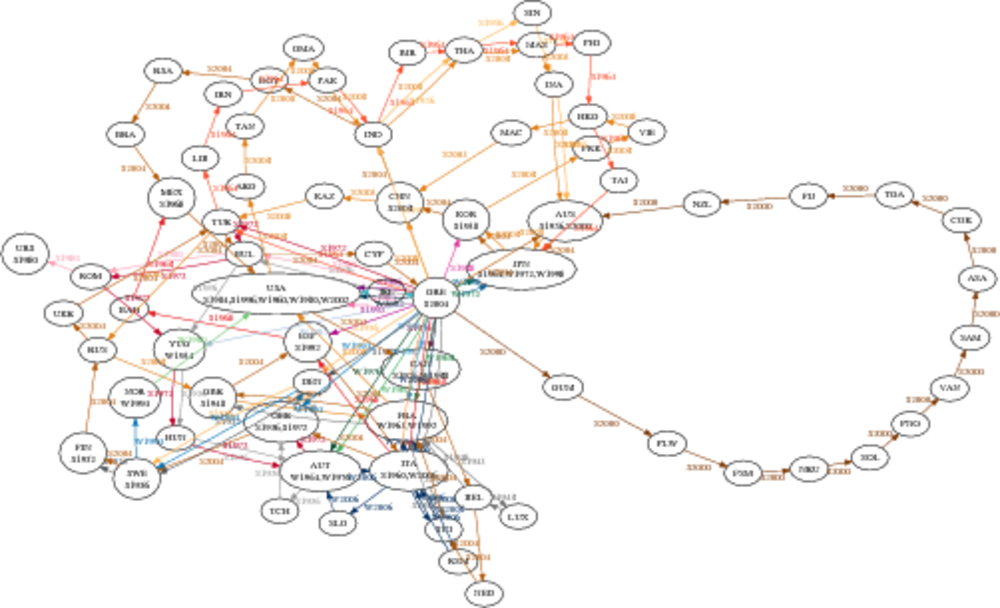}
  \caption{The Olympic torch relay graph with node overlaps removed, but before edge label overlaps are removed.\label{olympic_org}}
\end{center}
\end{figure}

Figure~\ref{olympic_org} shows the result of using PRISM on graph
representing the routes of Olympic torch relay from summer 1936 to
summer 2008. This graph is taken from the GD'08 contest.  Nodes represent
countries, and for a country that has hosted the Olympics, the year 
(e.g., ``S2008'' means summer 2008) that the Olympics was hosted is also included in the
node label. The edges are color codes, represented different
years and seasons the relays took place. Warm colors are used for summer games and cool color winter
games. Edge labels are used to highlight the year that edge was
traversed. In the figure, node overlaps are removed using
PRISM. However, due to the presence of edge labels, there are a lot of
overlaps among edge labels and between node and edge labels. As a
result, it is very difficult to follow the relay routes, even with the
help of the coloring scheme. This and other applications call for an
effective algorithm for removal node as well as edge label overlaps.

\subsection{The penalized proximity stress model for node and edge label overlap removal}

As Figure~\ref{olympic_org} demonstrates, when there are both node and
edge labels, simply removing node overlaps is not sufficient. Instead, we
need an algorithm to take into account the overlaps between both node
and edge labels.

The first algorithm we propose is a simple one. For each edge $\{i,\ j\}$
that has a label $k$, we break the edge down into two edges, $\{i,\ k\}$
and $\{k,\ j\}$, with the addition of a new node $k$. We call this expanded graph $G_a = \{V_a,\ E_a\}$. If we denote  $V_e$ the 
set of additional nodes derived from edge labels, then node set of the expanded graph
is a union of $V_e$ and the nodes in the original graph, or $V_a = V\cup V_e$.

\begin{figure}[htb]
  \centering
  \includegraphics[width=5.2in]{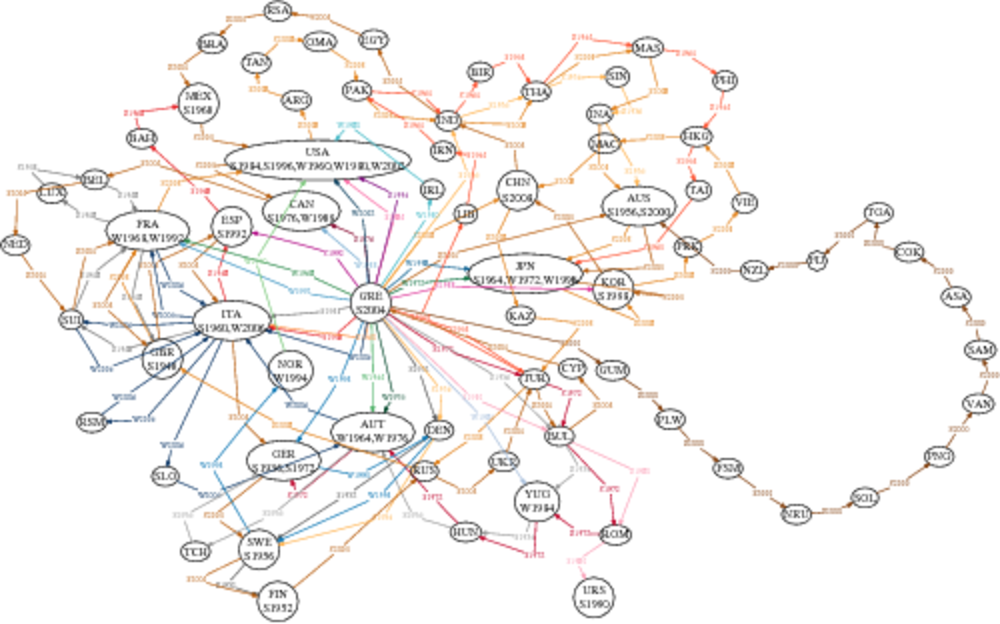}
  \caption{The Olympic torch relay graph after node and edge label overlaps are removed using vPRISM.\label{olympic_vlabel}}
\end{figure}

A simple algorithm for visualizing graphs with both node and edge 
labels is to apply PRISM to a layout of this expanded graph to
arrive at an overlap-free drawing. We call this algorithm
vPRISM. Figure~\ref{olympic_vlabel} gives the drawing of the Olympic
torch relay graph using vPRISM.  As can be seen, vPRISM is able to
avoid label overlaps and allows us to follow the routes of the
torch relay quite well. However, the shortcoming of this approach is also
evident. Because we are treating node labels and edge labels the same
way, some of the edges have very sharp bends. For example, at the far
left corner, the bend along the edge between ``LUX'' and ``BEL'', with
an edge label ``S1948'', is very sharp. While this may be alleviated
to some extent by the use of spline edges, it would be good to avoid
these bends in the first place.

We propose the following ePRISM algorithm. First, we lay out the graph 
without considering overlap removal. We can either lay out the
expanded graph $G_a$, resulting in a position for all node and edge labels,  or we can
lay out the original graph $G$, and place the edge labels at the center of the edges.
We then form a proximity graph through Delaunay triangulation using the position of both node and edge labels, and 
calculate the overlap factors along the edges of DT as in (\ref{ofactor}).
At this point, if we solve the proximity stress model (\ref{PRISM}), we will simply get
the vPRISM algorithm. Instead, we want to make sure that edges of the original graph
are as straight as possible. Suppose $k$ is a vertex represent the edge label
on edge $\{i,j\}\in E$, we want $x_k$ to be along the line between $x_i$ and $x_j$, 
preferably near the center of that line. Therefore we add to (\ref{PRISM}) a penalty term
$\left\|x_k - (x_i+x_j)/2\right\|^2$, leading to a {\em penalized proximity stress model}

\begin{equation}
\sum _{(i,j) \in  E_{P}} w_{\text{ij}}\left(\left\|x_i - x_j\| - d_{\text{ij}}\right.\right){}^2
+ \rho \sum_{k\in V_e, \{i,k\}\in E_a,  \{j,k\}\in E_a} w_k \left\|x_k-\frac{x_i+x_j}{2}\right\|^2\label{ePRISM}
\end{equation} 

\noindent to be minimized. Here $E_P$ is the edge set of the proximity graph of the expanded graph $G_a$, scalar $\rho$ is a penalty 
parameter, scaling factors $w_{\text{ij}} = 1/(d_{\text{ij}})^2$ are applied to the stress term, and
scaling factors $w_k = 1/||x_i-x_j||^2$, where $x_i$ and $x_j$ are the two nodes correspond to 
edge label $k$, are applied to the penalty term. 

The penalized proximity stress model (\ref{ePRISM}) can be solved using the stress majorization technique \cite{Gansner_Koren_North_2005b},
with a modification to account for the penalty term. This technique bounds the stress term by a sequence of
quadratic functions from the above. Finding the minimum of the quadratic functions can be formulated as 
the solution of sparse linear systems relating to the Laplacian of the 
proximity graph. In the case of our model, the linear system to be solved repeatedly is

\begin{equation}
\left(L_w + \rho L_e \right) x = L_{w,d}\ x^0 \label{sm}
\end{equation}

\noindent where $x^0$ is the current layout, and $x$ is the new layout that improves the stress and penalty terms.
The weighted Laplacian matrix $L_w$ has elements

\begin{equation*}
\left(L_w \right)_{\text{ij}}=\left\{ 
\begin{array}{ll}
 \sum _{\{i,l\}\in E_P} w_{\text{il}},& i = j \\
 -w_{\text{ij}}, & i \neq  j
\end{array}
 \right.
\end{equation*}

\noindent The matrix $L_e$ comes from penalty term and is defined as

\begin{equation*}
\left(L_e \right)_{\text{ij}}=\left\{ 
\begin{array}{ll}
  w_i,&\ i = j,\ i\in V_e\\
 0.25  \sum_{k\in N(i)\cap V_e} w_k ,&\ i = j,\ i\notin V_e \\
 -0.5  w_i,&\ i\in V_e,\ \{i,j\}\in E_a\\
 -0.5  w_j,&\ j\in V_e,\ \{i,j\}\in E_a\\
 0.25  w_k,&\ \{i,k\}\in E_a,\ \{j,k\}\in E_a,\ k\in V_e
\end{array}
 \right.
\end{equation*}

\noindent with $N(i)$ the set of neighboring nodes of $i$ in the expanded graph $G_a$, and matrix $L_{w,d}$ has elements

\begin{equation*}
\left(L_{w,d} \right)_{\text{ij}}=\left\{ 
\begin{array}{ll}
 \sum _{\{i,l\}\in E_P} w_{\text{il} }\left.d_{\text{il}}\right/\left\|y_i-y_l\right\|,& i = j \\
 -w_{\text{ij}} \left.d_{\text{ij}}\right/ \left\|y_i-y_j\right\|,& i \neq  j
\end{array}
 \right.
\end{equation*}

The addition of matrix $L_e$ to $L_w$ makes the overall matrix on the left hand side of 
(\ref{ePRISM}) somewhat denser, by introducing at most $6|E|$ extra entries if every edge has an edge label. However, as
long as the original graph is sparse, (\ref{sm}) is still a sparse system, and 
can be solved using conjugate gradient algorithm efficiently.
Algorithm~\ref{alg} gives a detailed description of the ePRISM algorithm.

\begin{algorithm}[H]
  \caption{ePRISM: an algorithm for node and edge label removal\label{alg}}
  \begin{algorithmic}
    \STATE Input: coordinates for each node and edge labels, $x^0_i$, and bounding box width and height 
$\{W_i, H_i\},\ i=1,2,\ldots,|V_a|.$
    \REPEAT
        \STATE  Form a proximity graph $G_{P}$ of $x^0$ by Delaunay triangulation.
        \STATE  Find the overlap factors (\ref{ofactor}) along all edges in  $G_{P}$. 
        \STATE Solve the penalized proximity stress model (\ref{ePRISM}) for $x$. Set $x^0 = x$.

     \UNTIL(convergence)
  
    \REPEAT
        \STATE  Form a proximity graph $G_{P}$ of $x^0$ by Delaunay triangulation.
        \STATE Find all node overlaps using a scan-line algorithm. Augment $G_{P}$ with edges from node pairs that overlap.
        \STATE  Find the overlap factor  (\ref{ofactor}) along all edges of $G_{P}$.
        \STATE Solve the proximity stress model (\ref{PRISM}) for $x$. Set $x^0 = x$.

     \UNTIL(convergence)
     
     \STATE Remove any remaining overlaps using PRISM.
  \end{algorithmic}
\end{algorithm}

Unlike PRISM, where the only objective is to remove overlaps, in
ePRISM, we have the dual objectives of removing overlaps as well as
keeping edges straight. Therefore, while in PRISM the lack of overlaps
is the measure of convergence, here, in the two main loops of the
ePRISM algorithm, we define convergence as $||x - x_0||/||x_0|| <
\epsilon$, with $\epsilon = 0.005$, and limit the number of iterations
to 1000. Because of the dual objectives, it is possible that after the two main loops have converged,
some overlaps may still exist, for example among edge
labels on multiple edges between the same pair of nodes. Therefore
we use PRISM algorithm to remove any overlaps that may still
remain. We set the penalty parameter $\rho$ to 4.

\begin{figure}[htp]
  \centering
  \includegraphics[width=5.2in]{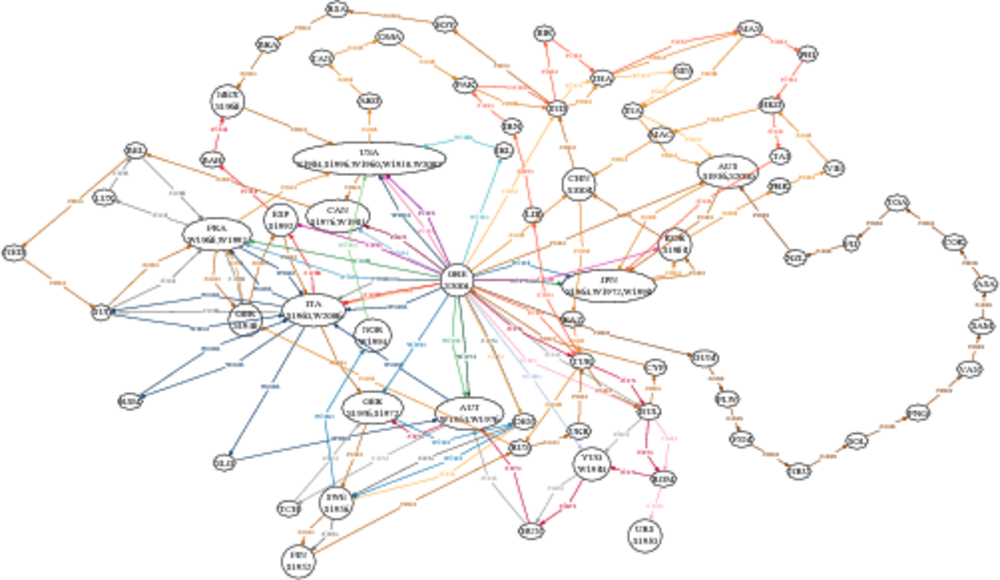}
  \caption{The Olympic torch relay graph after node and edge label overlaps are removed using ePRISM.\label{olympic_elabel}}
\end{figure}

The ePRISM algorithm has a similar computational complexity to
PRISM, when the latter is applied to the
expanded graph $G_a$. However we found that ePRISM tends to take more
iterations to converge in practice, due to the difficulty in
resolving the conflicting objective of removing label overlaps and
keeping edges straight.

Figure~\ref{olympic_elabel} gives the result of ePRISM on the Olympic
torch relay graph. Clearly, the majority of edges are now straight, and
those edges that are not straight have milder bends. Some
of these bends are unavoidable, such as when there are multiple
edges between the same pair of nodes.  In general, compared with
Figure~\ref{olympic_vlabel}, Figure~\ref{olympic_elabel} is more
pleasing to look at and easier to follow due to the great reduction of edge bends.

\begin{figure}[htb]
  \centering
  \includegraphics[width=5.2in]{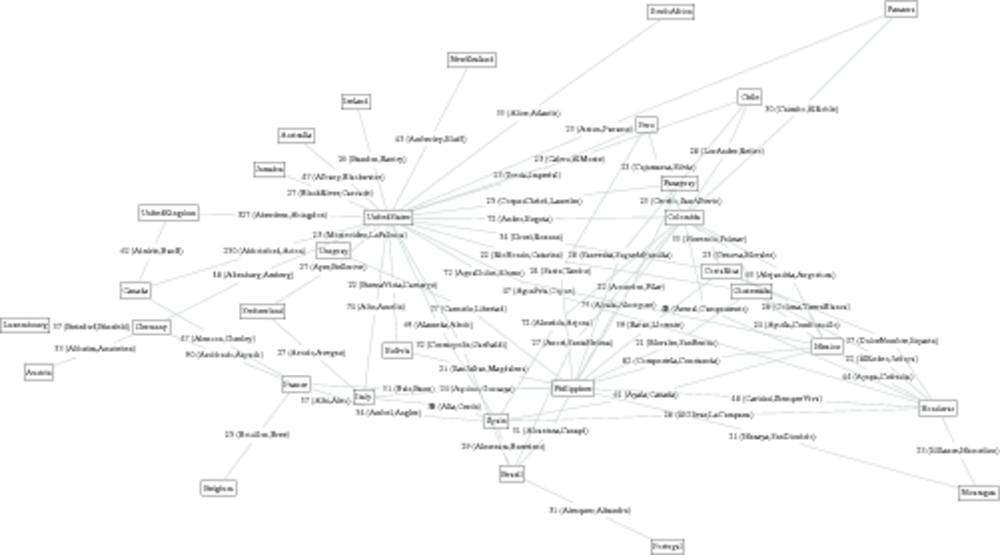}
  \caption{Visualization of countries sharing same city names. Two countries sharing more than 20 city names are linked by an edge, with the number of shared names, as well as two such names, shown as edge label. Node and edge label overlaps are removed using ePRISM.\label{country_city_elabel}}
\end{figure}

Finally, we apply ePRISM to another application. Here, we visualize countries and how they relate. Two countries are
related if there are more than 20 cities with the same name that exist in both countries. From the drawing, we can see that 
United States is the country with the most connections to other countries, reflecting the fact that most people in
the country emigrated from another country in the last three centuries, and that cities in the United States are often
named by the immigrants using names of cities in their home countries. The United States and the United Kingdom have the greatest number of cities with the same name (327),
followed by Canada (250), Germany (86), and Italy (74). The United States also shares city names with a large group of
countries in South America, seen on the right hand side of the figure. Canada, on the other hand, has the United States, the United Kingdom and
France as its three closest connected countries, reflecting the historical tie of Canada to these three countries.

\section{Conclusions and Future Work\label{sec_conclusion}}

The main contribution of this paper is a new algorithm for removing
both node and edge overlaps, based on a penalized proximity stress model.  The algorithm is shown to produce
layouts displaying both node and edge labels without overlaps, and is
aesthetic, with mostly straight edges.

For future work, we would like to investigate a combination of this algorithm
and edge label placement (ELP) techniques, so that those edge labels that can be placed
within the current geometry without introducing ambiguity or overlaps are dealt with separately to
those that have to be treated using the algorithm proposed in this paper.

\subsection*{\bf Acknowledgments}
The author would like to thank Emden Gansner and Stephen North for helpful discussions.

\bibliographystyle{titto-lncs-01.bst}
\bibliography{ref}

\end{document}